\documentclass[jkps,twocolumn,showpacs,floatfix]{revtex4}
\usepackage[dvips]{graphicx}
\begin{document}
\title{Empirical Formula Extended to the Yrast Excitation Energies \\of the Unnatural Parity States in Even-Even Nuclei}
\author{Dooyoung \surname{Kim}}
\author{Jin-Hee \surname{Yoon}}
\author{Dongwoo \surname{Cha}}
\email{dcha@inha.ac.kr}
\thanks{Fax: +82-32-866-2452}
\affiliation{Department of Physics, Inha University, Incheon
402-751, Korea}
\date{\today}

\begin{abstract}
Recently, it was shown that a simple empirical formula, in terms of the mass and valence nucleon numbers, can describe the main trends of the yrast excitation energies of the natural parity states up to $10^+$ in even-even nuclei throughout the entire periodic table. The same empirical formula was applied to the yrast excitation energies of unnatural parity states including $1^+$, $2^-$, $3^+$, $4^-$, $5^+$, $6^-$, $7^+$, $8^-$, $9^+$, $10^-$, and $11^+$ in even-even nuclei. Although the overall character of the effective residual interaction for the unnatural parity states was quite different to that of the natural parity states, the same form of the empirical formula was found to hold reasonably well for the yrast excitation energies of the unnatural parity states.
\end{abstract}


\maketitle

\section{Introduction}

We benefit sometimes by examining the particular nuclear properties in terms of simple nuclear variables over a wide span of the chart of nuclides. The oldest well known example is the Weizs{\" a}cker's semi-classical mass formula, which can reproduce the binding energy of the ground state of nuclei quite accurately in terms of the mass number $A$ and atomic number $Z$ \cite{Weizsacker}. Another well known study is the so-called $N_pN_n$ scheme, where $N_p$ and $N_n$ are the valence proton and neutron numbers, respectively. The $N_pN_n$ scheme denotes the phenomenon of a simple pattern that occurs when the nuclear data related to the lowest collective states are plotted against the product, $N_pN_n$. The $N_pN_n$ scheme has been used extensively and successfully for more than two decades to correlate the large volume of data on the collective degrees of freedom in nuclei \cite{Casten}.

Recently, another study devised an empirical formula by also adopting the valence nucleon numbers, to express the yrast excitation energies of the electric quadrupole ($E2$) states in even-even nuclei throughout the entire periodic table \cite{Ha}. Later, the same formula was shown to be capable of describing the main trends of the yrast excitation energies of not only the $E2$ states but also the natural parity, even multipole states up to $10^+$ found in all even-even nuclei \cite{Kim}. In addition, it was demonstrated that this empirical formula complied with the $N_pN_n$ scheme, even though the empirical formula itself did not explicitly depend on the product $N_pN_n$ \cite{Yoon}. Subsequently, the empirical formula was tested successfully for the yrast excitation energy of natural parity odd multipole states up to $9^-$ \cite{Jin}. Furthermore, it was also shown that the parameters fitted to each multipole can be represented by spin dependent parametrization with fewer parameters to be fitted with the data. Therefore one set of parameters can be applied to the yrast excitation energies of the entire natural parity even or odd multipole states in even-even nuclei \cite{Jin2}.

\begin{figure}[b]
\centering
\includegraphics[width=8.0cm,angle=0]{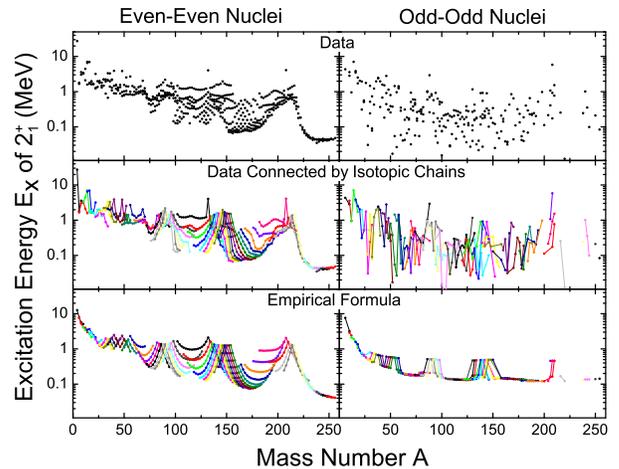}
\caption{The yrast excitation energy of the $E2$ states in even-even nuclei (left panel) and odd-odd nuclei (right panel). The measured excitation energies shown in the top two panels are connected along the isotopic chains in the middle panels. The calculated excitation energies by the empirical formula of Ref. \cite{Ha} are shown in the bottom two panels. The measured excitation energies were collected from the ENSDF database \cite{ENSDF}.}
\label{fig-1}
\end{figure}

It is quite remarkable to have a simple relationship in terms of the mass number and valence nucleon numbers between the measured yrast excitation energies of the given multipole states in all even-even nuclei. It is even more remarkable that the relationship has the same functional form regardless of the multipole of the excited states under consideration. However, it is better to discuss what it means to claim that there is a certain meaningful relationship between a myriad of data points. This issue is raised because the above mentioned empirical formula was sometimes critiqued its use of too many free parameters. This critique is certainly not unreasonable, considering John von Neumann's quote, ``With four parameters, I can fit an elephant, and with five I can make him wiggle his trunk \cite{Dyson}.''

As an example, this study considered the yrast excitation energies of the $E2$ states measured in even nuclei throughout the entire periodic table. Among these yrast excitation energies, those from the even-even nuclei are shown in the top-left panel of Figure\,\ref{fig-1}, while those from the odd-odd nuclei are shown in the top-right panel. The measured excitation energies were collected from the ENSDF database \cite{ENSDF}. Rather irregularly distributed data points can be seen in the top two panels. However, the same data points of the top left panel from the even-even nuclei appear to be arranged neatly after being connected along the isotopic chains, as shown in the middle left panel of Figure\,\ref{fig-1}. On the other hand, those of the top right panel from the odd-odd nuclei appear to be distributed irregularly even after being connected along the isotopic chains, as shown in the middle right panel. Finally, in the bottom panel of the same figure, the yrast excitation energies of the same $E2$ states as those of the upper two panels were calculated using the previously mentioned empirical formula introduced in Ref. \cite{Ha}. A comparison of the middle and bottom panels of Figure\,\ref{fig-1} shows that the measured yrast excitation energies of the even-even nuclei are represented well by the empirical formula, while those of the odd-odd nuclei are not.

The aim of this study was to collect all the yrast excitation energies of the unnatural parity states that had been measured up to now in even-even nuclei, and to determine if there is a simple relationship among them. We briefly introduce the empirical formula in Section II and then present the results on the yrast excitation energy of the unnatural parity states in even-even nuclei in Section III. Section IV presents the concluding remarks.

\section{Empirical Formula}

One does not normally attempt to describe any graph with many spikes, such as the one shown in the middle left panel of Figure\,\ref{fig-1}, using a simple formula, which depends on some variables that vary only monotonously. The empirical formula can represent such graphs because it employs the valence nucleon numbers $N_p$ and $N_n$. The valence proton (neutron) number $N_p$ ($N_n$) of a nucleus with an atomic (neutron) number $Z$ ($N$) is defined by
\begin{equation} \label{Val}
N_p (N_n) = \left\{ \begin{array}{ll}
   Z (N)-N_{c-1} & \mbox{if $N_{c-1} < Z (N) \leq  M_c$}\\
   N_c - Z (N)  & \mbox{if $M_c < Z (N) \leq N_c$}\end{array} \right.
\end{equation}
where $N_c$ is the magic number for the $c$-th major shell, which is given by $N_0 = 0$, $N_1= 8$, $N_2= 20$, $N_3= 28$, $N_4= 50$, $N_5= 82$, $N_6= 126$, etc. In addition, $M_c$ is the average of the two adjacent magic numbers, $(N_{c-1}+N_c)/2$, which corresponds to the number of nucleons contained in the mid-shell nucleus of the $c$-th major shell. The valence nucleon numbers, $N_p$ and $N_n$, are the  maximum for the mid-shell nuclei and zero for the magic shell nuclei. They repeat the positive integer numbers from zero whenever the atomic number $Z$ or neutron number $N$ crosses one of the major shell boundaries.

\begin{table}[b]
\begin{center}
\caption{Values adopted for the six parameters in Eq.\,(\ref{E6}) for the yrast excitation energy $E_x$ of the unnatural parity states in even-even nuclei. The last three columns are the $\chi^2$ value, standard deviation $\sigma$ and total number $N_0$ of data points, respectively, for the corresponding multipole state.}
\begin{tabular}{crccccccc}
\hline\hline
~~$J_1^\pi$~~&~~$\alpha$~~&~~$\gamma$~~&~~$\beta_p$($\beta_n$)~~
&~~$\lambda_p$( $\lambda_n$)~~&~~$\chi^2$~~&~~$\sigma$~~&~~$N_0$~~\cr
&(MeV)&&(MeV)&&&&\cr
\hline
$1_1^+$&47.13&0.67&0.54(0.99)&0.76(0.50)&0.079&0.28&251\cr
$3_1^+$&49.46&0.76&1.17(1.49)&0.58(0.32)&0.051&0.22&236\cr
$5_1^+$&87.00&0.82&1.05(1.26)&0.40(0.24)&0.028&0.16&250\cr
$7_1^+$&139.02&0.88&1.19(1.48)&0.28(0.24)&0.021&0.14&184\cr
$9_1^+$&172.81&0.86&1.09(1.61)&0.34(0.46)&0.019&0.13&159\cr
$11_1^+$&350.33&0.97&0.90(1.93)&0.12(0.38)&0.018&0.13&117\cr
\hline
$2_1^-$&48.27&0.73&1.09(1.59)&0.19(0.31)&0.058&0.23&246\cr
$4_1^-$&75.04&0.81&1.00(1.27)&0.17(0.24)&0.027&0.16&253\cr
$6_1^-$&107.89&0.83&0.77(1.40)&0.19(0.28)&0.019&0.13&248\cr
$8_1^-$&277.43&1.00&0.90(1.49)&0.15(0.20)&0.017&0.13&230\cr
$10_1^-$&238.48&0.90&1.24(1.76)&0.44(0.25)&0.013&0.11&199\cr
\hline\hline
\end{tabular}
\label{tab-1}
\end{center}
\end{table}

The original four parameter form of the empirical formula first introduced in Ref.\,\cite{Ha} for the yrast excitation energy of the $E2$ states in even-even nuclei is written as follows:
\begin{equation} \label{E4}
E_x = \alpha A^{-\gamma} + \beta \left[ e^{- \lambda   N_p } + e^{- \lambda   N_n } \right],
\end{equation}
where $\alpha$, $\gamma$, $\beta$ and $\lambda$ are the model parameters to be fitted from the data. However, after testing different formulae with several other forms including a term with the product $N_p N_n$, the following six parameter form,
\begin{equation} \label{E6}
E_x  = \alpha  A^{-\gamma}  + \beta_p  e^{- \lambda_p   N_p }  + \beta_n e^{- \lambda_n N_n }
\end{equation}
were chosen as the most appropriate expression for the yrast excitation energy of the $E2$ states in even-even nuclei \cite{Jin3}. Here, the parameters, $\beta$ and $\lambda$, in Eq.\,(\ref{E4}) are split into $\beta_p$, $\beta_n$ and $\lambda_p$, $\lambda_n$, respectively. This considers the fact that protons and neutrons make different contributions to the yrast excitation energy, $E_x$.

\begin{figure*}[t]
\centering
\includegraphics[width=14.0cm,angle=0]{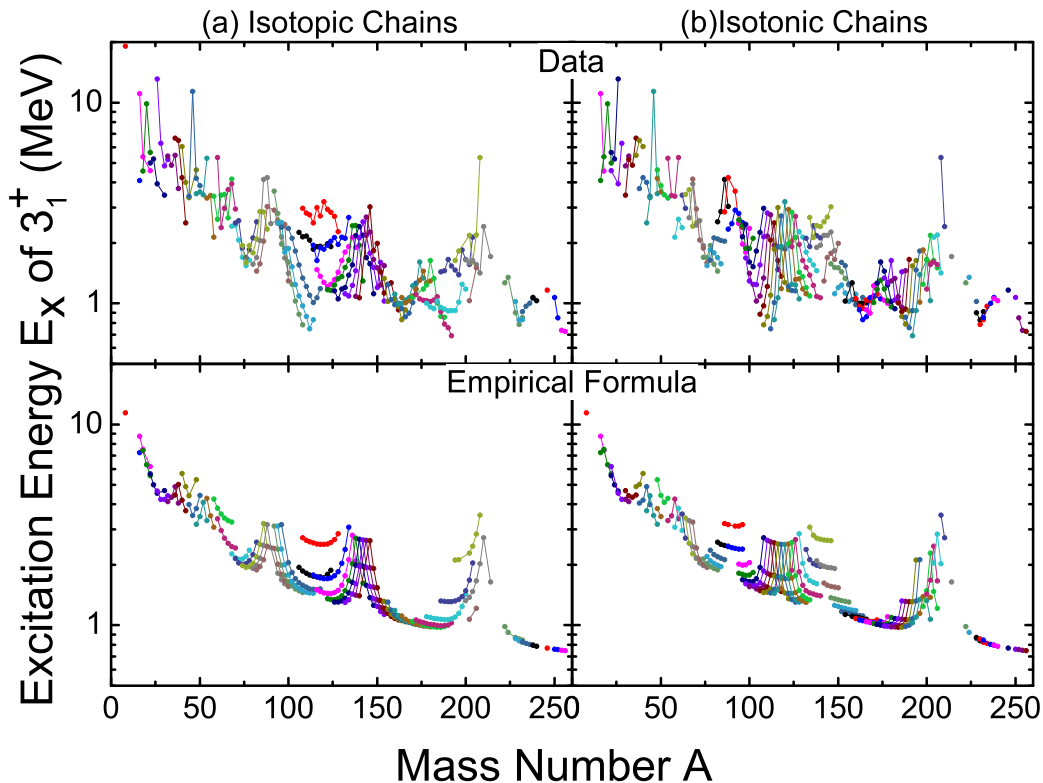}
\caption{The yrast excitation energies of $M3$ states in even-even nuclei. The data points are connected by solid lines along the isotopic chains in (a) and along the isotonic chains in (b). The upper two panels show the measured yrast excitation energies while the lower two panels show those calculated by the empirical formula given by Eq.\,(\ref{E6}). The measured excitation energies were collected from the ENSDF database \cite{ENSDF}.}
\label{fig-2}
\end{figure*}

The six model parameters were fixed using the usual least-squares-fitting (LSF) technique. However, in the LSF process, instead of taking the usual error, $E_x^{\rm cal} (i) - E_x^{\rm exp} (i)$, between the calculated yrast excitation energy $E_x^{\rm cal} (i)$ and measured one, $E_x^{\rm exp} (i)$, for the $i$-th data point, one should employ the logarithmic error $R_E(i)$ defined by
\begin{equation} \label{Re}
R_E(i) =  \log \big[ E_x^{\rm cal}(i) \big] - \log \big[ E_x^{\rm exp}(i) \big]
\end{equation}
in order to consider the fact that the values of $E_x$ span a broad range differing by as much as an order of two \cite{Sabbey}. The values of the model parameters were determined by minimizing the $\chi^2$ value, which was defined in terms of the logarithmic error $R_E (i)$ by
\begin{equation} \label{Chi}
\chi^2 = { 1 \over {N_0 - p}} \sum_{i=1}^{N_0} \Big| R_E(i) \Big|^2
\end{equation}
where $N_0$ and $p$ are the number of total data points considered and the number of model parameters, respectively. The LSF process was carried out for each multipole and different minimum $\chi^2$ values were obtained for different multipoles. However, the $\chi^2$ value itself is not a good measure for comparing the degrees of fitting for different multipoles. Instead, this study adopted the usual standard deviation $\sigma$ defined by
\begin{equation} \label{Sigma}
\sigma = \sqrt{ { 1 \over N_0} \sum_{i=1}^{N_0} \Big| R_E(i) \Big|^2 }
\end{equation}
which is practically the same as the square root of the $\chi^2$ value when $N_0 \gg p$.

\section{Yrast Excitation Energies of Unnatural Parity States in Even-even Nuclei}

For the last two years, it has been shown that Eq.\,(\ref{E6}) can be used to describe the entire set of the yrast excitation energies measured in all the even-even nuclei for the natural parity states, including even multipoles up to $10^+$ as well as odd multipoles up to $9^-$ \cite{Kim,Jin}. Motivated by such results on the natural parity states, this study collected all the measured yrast excitation energies of the unnatural parity states in even-even nuclei from the ENSDF database \cite{ENSDF}. The model parameters $\alpha$, $\gamma$, $\beta_p$, $\beta_n$, $\lambda_p$ and $\lambda_n$ of Eq.\,(\ref{E6}) were determined using the LSF process for each multipole state, including the odd multipole states up to $11^+$ and even multipole states up to $10^-$. Table\,\ref{tab-1} lists the fitted parameter values with the $\chi^2$ value, standard deviation $\sigma$ and total number of data points $N_0$ of the corresponding multipole state. These results, when compared to those of the natural parity states, show that Eq.\,(\ref{E6}) can describe the yrast excitation energies of the unnatural and natural parity states. The parameter values of this study were determined similarly to those of the natural parity case, and the standard deviation $\sigma$ of the unnatural parity multipoles were the same or even less than those obtained for the natural parity multipoles.

Indeed, Eq.\,(\ref{E6}) was suitable for both the natural and unnatural parity states. It has been known for quite a long time \cite{Speth} that, for the natural parity states, the lowest excited state is usually a collective one because the spin-isospin independent effective residual interaction is mainly attractive . Therefore, among the given natural parity multipole states, the lowest yrast state can represent the main characteristics of that multipole. However, the spin-isospin dependent effective residual interaction, which governs the collectivity of the unnatural parity states, is mainly repulsive. That means the collective states among the given unnatural parity multipole states lie at a somewhat higher part of the excitation energy spectrum and the yrast excitation energy can be thought to be determined quite irregularly. Therefore, it is difficult to imagine that, in the case of the unnatural parity states, the yrast excitation energies, whose values are in some sense chosen arbitrarily, will follow a given routine.

\begin{figure}[t]
\centering
\includegraphics[width=8.0cm,angle=0]{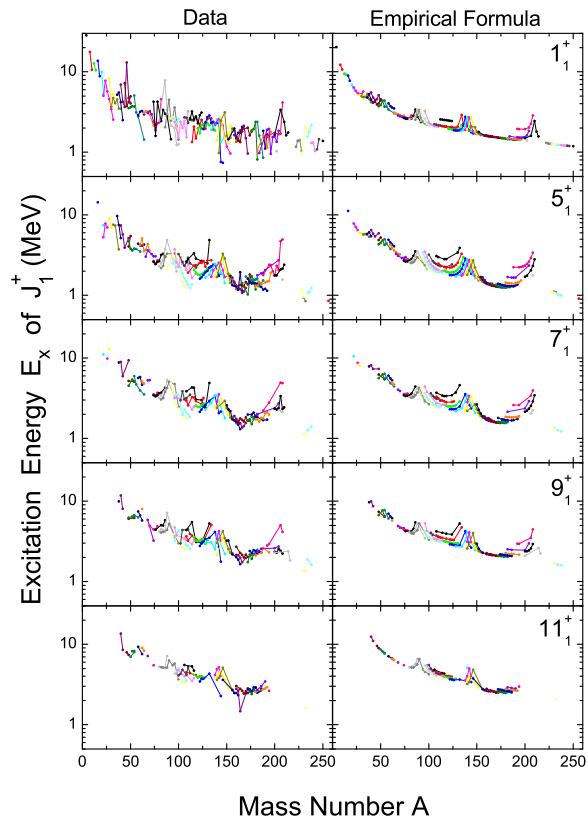}
\caption{The yrast excitation energy of the unnatural parity odd multipole states including $1^+$, $5^+$, $1^+$, $9^+$, and $11^+$ in even-even nuclei. The measured excitation energies and those calculated by Eq.\,(\ref{E6}) are shown in the left and right panels, respectively. All data points are connected along the isotopic chains. The measured excitation energies were collected from the ENSDF database \cite{ENSDF}.}
\label{fig-3}
\end{figure}

\begin{figure}[t]
\centering
\includegraphics[width=8.0cm,angle=0]{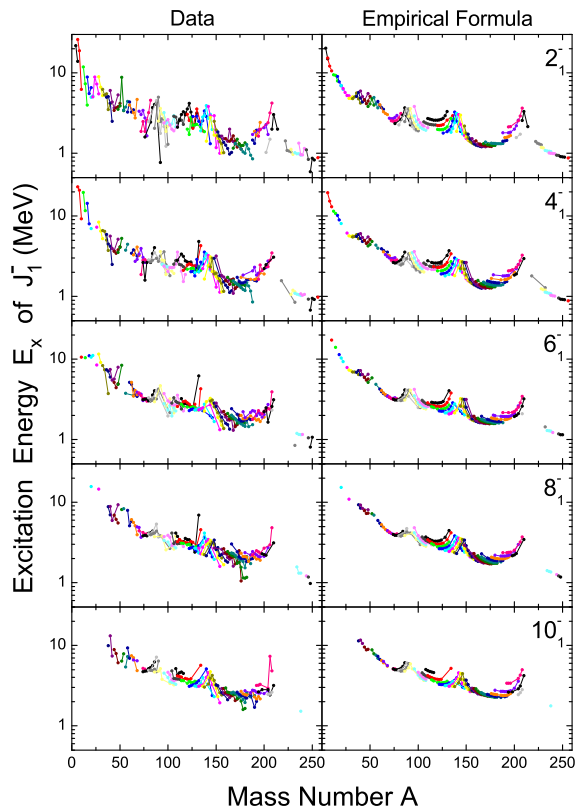}
\caption{Same as in Figure\,\ref{fig-3} but for the unnatural parity even multipole states including $2^-$, $4^-$, $6^-$, $8^-$ and $10^-$ in even-even nuclei. The measured excitation energies were also collected from the ENSDF database \cite{ENSDF}.}
\label{fig-4}
\end{figure}

Three graphs were prepared (Figures\,\ref{fig-2}-\ref{fig-4}) to determine how well Eq.\,(\ref{E6}) describes the yrast energies of the unnatural parity states, where the data was compared with the yrast excitation energies calculated by Eq.\,(\ref{E6}). We begin by picking up one typical multipole state among the eleven unnatural parity multipoles, namely the magnetic octupole ($M3$), and showing their yrast excitation energies in more detail (Figure\,\ref{fig-2}). The upper panels of the figure show the measured yrast excitation energies while the lower panels of the same figure show those energies calculated using Eq.\,(\ref{E6}), using the parameter listed in Table\,\ref{tab-1}. The data points in Figure\,\ref{fig-2}(a) are connected along the isotopic chains while those in Figure\,\ref{fig-2}(b) are connected along the isotonic chains. (In the electronic version, the color code for the isotopic (isotonic) chains in Figure\,\ref{fig-2}(a) is the same as the color code for the corresponding isotopic (isotonic) chains in Figure\,\ref{fig-2}(b).) By comparing the measured yrast excitation energies (upper two panels) with the calculated ones (lower two panels), it can be seen that Eq.\,(\ref{E6}) can explain the essential trends of the $M3$ yrast excitation energies measured in even-even nuclei throughout the entire periodic table. However, we do not claim that each one of the calculated yrast excitation energies represents a theoretical estimate of the yrast excitation energy of the corresponding state. Rather, there is a certain relationship between the yrast excitation energies of the same unnatural parity multipole, like $M3$, which are observed in different even-even nuclei. Moreover, the empirical formula, Eq.\,(\ref{E6}), can reveal the main characteristics of that relationship. For example, the structure of the $A$ dependent shape due to major shell closure is clearly reproduced and the distinction between the shapes of the isotopic and the isotonic chains is also demonstrated by Eq.\,(\ref{E6}). However, some regions where Eq.\,(\ref{E6}) cannot properly reproduce the obvious bulged downward structure are shown by the data at $A \approx 110$, $165$, $190$ and $230$. However, this type of discrepancy was also observed in the case of the natural parity odd-multipole states, which was attributed to some strong multipole correlations between nucleons at the Fermi level \cite{Jin}.

The performance of Eq.\,(\ref{E6}) on the yrast excitation energies of the unnatural parity states other than the $M3$ state can be examined by inspecting the next two figures. The yrast excitation energies of the unnatural parity odd multipole states up to $11^+$ were plotted as a function of the mass number $A$ in Figure \ref{fig-3}. The yrast excitation energies of the unnatural parity even multipole states up to $10^-$ were plotted as a function of the mass number $A$ in Figure \ref{fig-4}. The left panels of these two figures show the measured yrast excitation energies while the right panels of the same figures show those energies calculated by Eq.\,(\ref{E6}) using the parameter set listed in Table\,\ref{tab-1}. Both the measured and calculated data points are connected along the isotopic chains. (In the electronic version, the data points that belong to the same isotopic chain are shown by the same colored symbols, and the color code for the measured data points in the left panels is the same as the color code for the corresponding calculated points in the right panels.) Observing these graphs, it can be seen that the overall $A$ dependent shape of the data points was reproduced well for all the multipoles considered in Figures\,\ref{fig-3} and \ref{fig-4} in a similar manner to the case of the $M3$ state shown in Figure\,\ref{fig-2}. In addition, a comparison of these graphs with those of similar studies on the natural parity states from Refs. \cite{Kim} and \cite{Jin}, showed that the results on the unnatural parity states were comparable to those on the natural parity states.

\begin{figure}[t]
\centering
\includegraphics[width=8.0cm,angle=0]{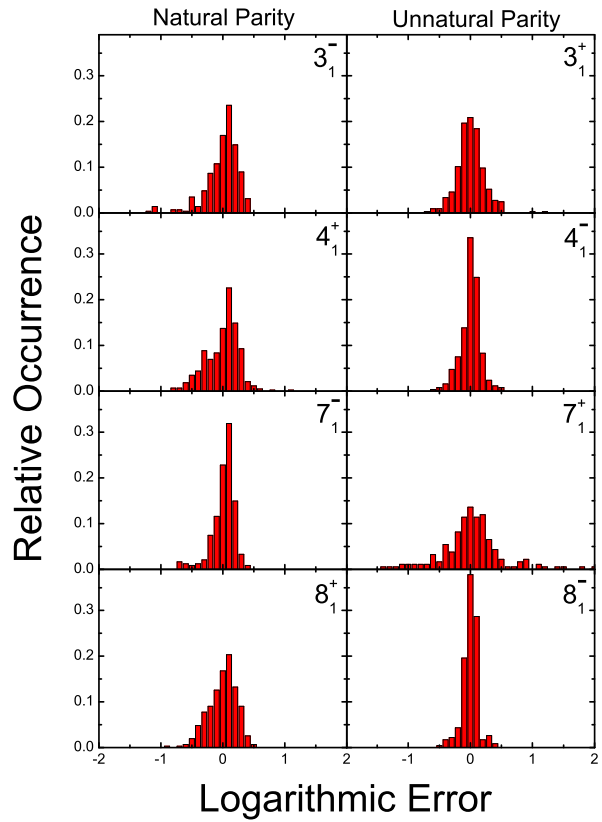}
\caption{The histograms for the occurrence of the logarithmic error $R_E$ defined by Eq.\,(\ref{Re}) for four randomly chosen natural (left panels) versus unnatural (right panels) parity multipole states. The area under each histogram was normalized to unity.}
\label{fig-5}
\end{figure}

Finally, this study compared the histograms for the occurrence of the logarithmic error $R_E$ defined by Eq.\,(\ref{Re}) in Figure\,\ref{fig-5} for four randomly chosen natural (left panels) versus unnatural (right panels) parity multipole states. The abscissa and ordinate of Figure\,\ref{fig-5} denote the logarithmic error and relative number of occurrence, respectively. The area under each histogram (the total relative number of occurrence) was normalized to unity in order to make a proper comparison of the degree of the fit between different multipoles. A comparison of the histograms of the unnatural parity states with those of the natural parity states from Figure\,\ref{fig-5} shows that the performance of Eq.\,(\ref{E6}) for the unnatural parity states is as good as, if not better, than that for the natural parity states. Such an assessment is reinforced when the standard deviations listed in Table\,\ref{tab-1}, such as $0.22(3^+)$, $0.16(4^-)$, $0.14(7^+)$, and $0.13(8^-)$ for the unnatural parity states, are compared with those quoted from Refs. \cite{Kim} and \cite{Jin}, such as $0.27(3^-)$, $0.27(4^+)$, $0.19(7^-)$, and $0.23(8^+)$ for the natural parity states.

\section{Concluding Remarks}

These results show that the empirical formula, Eq.\,(\ref{E6}), can be applied reasonably well to describe the essential trends of the yrast excitation energies of the unnatural parity states in even-even nuclei throughout the entire periodic table. This work completes a series of studies where the empirical formula was tested on the yrast excitation energies of various different multipole states. It started with the $E2$ states where the measured yrast excitation energies, when plotted against the mass number $A$, show an already distinctive pattern, as shown in Figure\,\ref{fig-1} \cite{Ha}. Although the agreement between the overall shapes made by the measured $E2$ yrast excitation energies and the calculated ones were impressive, the empirical formula itself did not gain much recognition because it was obtained rather accidentally and the physical origin of each term in the empirical formula could not be obtained, except for a couple of discussions regarding the possible interpretation \cite{Ha2,Ha3}.

 Nevertheless, the test of the empirical formula on the yrast excitation energies was carried out first to the higher even multipole natural parity states, such as $4^+$, $6^+$, $8^+$, and $10^+$, and then to the natural parity odd multipole states, including $1^-$, $3^-$, $5^-$, $7^-$, and $9^-$. Such tests always ended up with affirmative results as already explained. Even the yrast excitation energies of the unnatural parity states followed the empirical formula, even though such results were unexpected. Therefore, the empirical formula, Eq.\,(\ref{E6}), characterize the overall shape of the yrast excitation energy of all multipoles including both the natural and unnatural parity states.

 On the other hand, this conclusion strongly suggests that there is a certain relationship between the yrast excitation energies regardless of their multipole or parity. It should be emphasized that it is the empirical formula, Eq.\,(\ref{E6}), even though the empirical formula itself does not attract much recognition, which allows us to comprehend the very existence of a relationship between the yrast excitation energies. Once it has been established that there is a universal relationship between the yrast excitation energies, it is natural to imagine that there would some underlying dynamical origin. Unfortunately, the origin is unclear. However, the functional form of the empirical formula, Eq.\,(\ref{E6}), might provide a clue. Therefore, further study will be needed.

\begin{acknowledgments}
This work was supported by an Inha University Research Grant.
\end{acknowledgments}

\end{document}